\documentclass[12pt]{article}

\ifx\pdfoutput\undefined
\usepackage[dvips,bookmarks]{hyperref}
\else
\usepackage{hyperref}
\fi
\hypersetup{colorlinks=false,bookmarksopen,bookmarksnumbered,citecolor=blue,
   pdfstartview=FitH}

\usepackage{amssymb}
\usepackage{makeidx}
\usepackage{graphicx}
\usepackage{ccaption}
\usepackage{subfigure}
\usepackage{rotating}
\usepackage{lscape}
\usepackage{verbatim}
\usepackage{amsmath, amsthm,amssymb}
\usepackage{mathrsfs}
\usepackage{hypernat}
\usepackage{hyperref}
\usepackage{amsfonts}
\usepackage{dsfont}
\usepackage{bbm}
\usepackage{slashed, tensor}

\usepackage{pdfpages}

\usepackage{amsopn}

\parskip=\medskipamount

\usepackage[mathscr]{euscript} 

\usepackage{amsmath}
\usepackage{amsfonts}
\usepackage{slashed}
\usepackage{youngtab}

\usepackage{breqn}
\usepackage{booktabs}

\usepackage{tikz}
\usetikzlibrary{arrows,shapes}

\usepackage{hyperref}

\newcommand {\cD}{{\cal D}}
\newcommand {\cE}{{\cal E}}

\newcommand {\cN}{{\cal N}}

\newcommand {\cU}{{\cal U}}
\newcommand {\cV}{{\cal V}}

\newcommand {\cX}{{\cal X}}
\newcommand {\cY}{{\cal Y}}

\def\a{\alpha}
\def\b{\beta}

\def\d{\delta}

\def\f{\phi}

\def\k{\kappa}
\def\l{\lambda}
\def\m{\mu}

\def\q{\theta}

\def\s{\sigma}
\def\t{\tau}

\def\F{\Phi}
\def\J{\Psi}

\def\S{\Sigma}

\def\X{\Xi}

\def\ri{{\rm i}}
\def\re{{\rm e}}

\newcommand{\ad}{{\dot{\alpha}}}
\newcommand{\bd}{{\dot{\beta}}}

\newcommand{\pa}{\partial}
\newcommand{\hf}{\frac12}

\newcommand{\be}{\begin{equation}}
\newcommand{\ee}{\end{equation}}
\newcommand{\bea}{\begin{eqnarray}}
\newcommand{\eea}{\end{eqnarray}}
\newcommand{\non}{\nonumber}
\newcommand{\ba}{\begin{array}}
\newcommand{\ea}{\end{array}}

\newcommand{\bm}[1]{\mbox{\boldmath$#1$}}

\def\double #1{#1{\hbox{\kern-2pt $#1$}}}

\newcommand{\bsubeq}{\begin{subequations}}
\newcommand{\esubeq}{\end{subequations}}

\newcommand{\rd}{\mathrm d}
\numberwithin{equation}{section}  

\begin{document}

\begin{titlepage}
\begin{flushright}
May, 2017 \\
\end{flushright}
\vspace{5mm}

\begin{center}
{\Large \bf Three-form multiplet and supersymmetry breaking}
\\ 
\end{center}

\begin{center}

{\bf Evgeny I. Buchbinder and Sergei M. Kuzenko} \\
\vspace{5mm}

\footnotesize{
${}^{a}${\it School of Physics and Astrophysics M013, The University of Western Australia\\
35 Stirling Highway, Crawley W.A. 6009, Australia}}  
\vspace{2mm}
~\\
\texttt{ evgeny.buchbinder@uwa.edu.au, sergei.kuzenko@uwa.edu.au
}\\
\vspace{2mm}

\end{center}

\begin{abstract}
\baselineskip=14pt
Recently, a nilpotent real scalar superfield $V$ was introduced in 
 \cite{KMcAT-M} as a model for the Goldstino. It contains only two 
independent component fields, the Goldstino and the auxiliary $D$-field.  
Here we first show that $V$ can equivalently be realised as 
 a constrained three-form superfield.
We demonstrate that every irreducible Goldstino superfield 
(of which  the Goldstino is the only independent component field) 
can be realised as a descendant of $V$ which is invariant under local rescalings 
$V  \to  \re^\t V$, where $\tau$ is an arbitrary real scalar superfield. 
We next propose a new Goldstino supermultiplet which is  described by a
nilpotent three-form superfield $\cY$ that is 
 a variant formulation for the nilpotent chiral superfield, 
which is often used in off-shell models for spontaneously broken supergravity. 
It is shown that the action describing the dynamics of  $\cY$ 
may be obtained from a supersymmetric nonlinear $\s$-model in  the infrared limit. 
Unlike $V$, the Goldstino superfield $\cY$ contains two independent
auxiliary fields, $F= H+\ri G$, 
of which $H$  is a scalar and $G$ is the field strength of a gauge three-form. 
When $\cY$ is coupled to supergravity, both $H$ and $G$ produce positive contributions to the cosmological constant.  While the contribution from $H$ is uniquely determined by the parameter of the supersymmetry breaking in the action, 
the contribution from $G$ is dynamical. 
\end{abstract}

\vfill

\vfill
\end{titlepage}

%
%

\renewcommand{\thefootnote}{\arabic{footnote}}
\setcounter{footnote}{0}


\allowdisplaybreaks

\section{Introduction}
\label{section1}

Recently, a new Goldstino superfield for spontaneously broken $\cN=1$  
supersymmetry in four spacetime dimensions has been introduced \cite{KMcAT-M}. 
In the super-Poincar\'e case, it is described by a real scalar superfield
 $V$ subject to the nilpotency constraints 
\begin{subequations} \label{1.1}
\bea
V^2&=&0~, \\
V D_A D_B V &=&0~, \label{1.1b}\\
V D_A D_B D_C V &=&0~, \label{1.1c}
\eea
\end{subequations}
where $D_A =(\pa_a , D_\a, \bar D^\ad)$ are the covariant derivatives of 
$\cN=1$ Minkowski superspace ${\mathbb M}^{4|4}$.
These nilpotency constraints have to be supplemented 
with the requirement that  the descendant 
$D^\a W_\a =\bar D_\ad \bar W^\ad$ is nowhere vanishing, where
\bea
W_\a = -\frac{1}{4} \bar D^2  D_\a V~.
\label{1.2}
\eea
The nilpotency constraints \eqref{1.1} imply that $V$ has two independent component fields\footnote{The component analysis of $V$ simplifies 
if one notices that the constraints 
\eqref{1.1b} and \eqref{1.1c} are equivalent to  $V\pa_A \pa_B V=0$ 
and $V \pa_A \pa_B \pa_C V=0$, respectively.}
which are:  the Goldstino, which may be identified with  $W_\a|_{\q=0}$, and  the auxiliary $D$-field  being proportional to $D^\a W_\a|_{\q=0}$. All other component fields of $V$ are composite ones. In particular,  from \eqref{1.1}  
one may deduce the representation 
\bea
V = - 4 \frac{W^2 \bar W^2}{(D^\a W_\a)^3}~, \qquad 
W^2 = W^\a W_\a~,
\label{1.3}
\eea
which turns  \eqref{1.1} into identities.
The dynamics of this supermultiplet is governed by the action
\bea
S =   \int \rd^4 x \rd^2 \q  \rd^2 \bar{\q} \,\Big\{
\frac{1}{16} V D^\a \bar D^2 D_\a V- 2f  V\Big\}~,
\label{1.4}
\eea
where $f$ is a positive parameter of mass dimension $+2$ which characterises 
the supersymmetry breaking scale. 

The constraints \eqref{1.1} are invariant under local re-scalings of $V$,
\bea
V ~ \to ~ \re^\t V~,
\label{1.5}
\eea
where $\t$ is an arbitrary real scalar superfield. Requiring the action \eqref{1.4} to be stationary under arbitrary variations of the form $\d V = \t V$
leads to the following constraint 
\bea
f  V =\frac{1}{16} V D^\a \bar D^2 D_\a V~,
\label{1.6}
\eea
which proves to express the auxiliary field in terms of the Goldstino.
The constraints \eqref{1.1} and \eqref{1.6} define the irreducible 
Goldstino superfield $\cV$ described in \cite{BHKMS}.

An explicit realisation for $\cV$ 
was given long ago by Lindstr\"om and Ro\v{c}ek \cite{LR}
in the form 
\bea
f \cV = \bar \f \f~, 
\eea
where $\f$ is the irreducible chiral scalar Goldstino superfield 
\cite{Rocek,IK78},
$\bar D_{\dot\alpha} \f =0$, which is
subject to the constraints \cite{Rocek}: 
\begin{subequations} 
\label{1.8}
\bea 
\f^2&=&0 ~,  \label{1.8a}\\
{f} \f &=& -\frac 14  \f {\bar D}^2 \bar  \f ~.
\label{1.8b}
\eea
\end{subequations}
Another useful realisation for $\cV$ is 
\bea
f \cV = \bar \S \S~, 
\label{1.9}
\eea
where $\S$ is the 
Goldstino superfield introduced in
\cite{KTyler}. It obeys the improved complex linear constraint
\begin{subequations}\label{cls}
\bea
-\frac 14 {\bar D}^2\Sigma={f}\,,
\eea
as well as it is nilpotent and is subject to a holomorphic nonlinear constraint, 
\bea
\S^2 &=&0\, ,  \label{1.10b} \\
f D_\alpha\Sigma &=& 
-\frac 14 \Sigma{\bar D}^2D_\alpha\Sigma
~.
\eea
\end{subequations}

In this paper we uncover new interesting properties of the Goldstino multiplet $V$.
In particular, we show that $V$ can equivalently be realised as 
 a constrained three-form superfield. We also demonstrate that every irreducible Goldstino superfield\footnote{The notion of irreducible and reducible Goldstino
 superfields was introduced in \cite{BHKMS}, see also section \ref{section3} below.}  
 can be realised as a descendant of $V$, which is invariant under local rescalings 
$V  \to  \re^\t V$, where $\tau$ is an arbitrary real scalar superfield. 

In the last two years, there has been much interest in models for off-shell $\cN=1$ supergravity coupled to nilpotent Goldstino superfields. One of the main reasons for such interest 
is that 
 every Goldstino superfield coupled to supergravity provides a positive contribution to the cosmological constant 
\cite{LR,DFKS,BFKVP,HY,K15,BHKMS}, unlike the supersymmetric cosmological term 
 \cite{Townsend} which yields
a negative contribution to the cosmological constant. 
The same property holds for the Goldstino brane \cite{BMST}.
All irreducible Goldstino superfields provide one and the same 
universal contribution 
to the cosmological constant \cite{BHKMS}, 
in accordance with the super-Higgs effect \cite{VS,VS2,DZ}.
Exactly the same contribution is generated by the known reducible Goldstino 
multiplets, which are: (i) the superfield $V$ discussed above; and 
(ii) the nilpotent chiral superfield $\cX$ \cite{Casalbuoni,KS}
used in numerous recent publications including 
\cite{DFKS,BFKVP,HY}.  In this paper paper we introduce a new reducible 
Goldstino superfield, a nilpotent three-form multiplet, which will be shown 
to provide two 
separate positive contributions to the cosmological constant, of which one is universal 
and the other is dynamical.


\section{From $V$ to equivalent three-form multiplet}
\label{section2}

We start by recalling two simple models for spontaneously broken supersymmetry
that are realised in terms of an {\it unconstrained}  real scalar superfield $V$. 
One of them describes an abelian  vector multiplet with action
\bea
S_{\rm VM} = \Big\{ \frac{1}{8}  \int \rd^4 x \rd^2 \q \, W^\a W_\a +{\rm c.c.} \Big\}
- 2f  \int \rd^4 x \rd^2 \q  \rd^2 \bar{\q} \, V~,
\label{2.1}
\eea
which is obtained by adding 
a Fayet-Iliopoulos term \cite{FI} to the massless vector multiplet action.
The functional \eqref{2.1}  is invariant under  gauge transformations
\bea
\d V= \l +\bar \l~, \qquad \bar D_\ad \l =0~.
\label{2.2}
\eea
The chiral spinor $W_\a$ defined by \eqref{1.2}
is a gauge-invariant field strength for this model.

Our second model describes the dynamics of a variant scalar multiplet known as the three-form multiplet\cite{Gates,GS} (see \cite{GGRS} for a review).
We choose the action 
\bea
S_{\rm TFM} = \int \rd^4 x \rd^2 \q  \rd^2 \bar{\q} \, \bar \J \J 
-  \Big\{f  \int \rd^4 x \rd^2 \q \, \J + {\rm c.c.} \Big\}~,
\label{2.3}
\eea
where the chiral scalar superfield $\J$, $  \bar D_\ad \J =0$, is defined by 
\bea
\J = -\frac{1}{4} \bar D^2 V ~.
\label{2.4}
\eea
It is called the three-form multiplet.
Its  specific feature is the relation
\bea
D^2 \J - \bar D^2 \bar \J =  \ri \pa^{\a \ad} v_{\a\ad} ~, \qquad 
v_{\a\ad} = [D_\a , \bar D_\ad ] V~,
\label{2.5}
\eea
which means that the auxiliary $F$-field of $\J$, 
defined  by $F=-\frac{1}{4} D^2 \J |_{\q=0}$, is a complex scalar such that 
its imaginary part,  ${\rm Im} F$,  is the divergence of a vector 
(or, equivalently, the field strength of a gauge three-form).
This is the only difference between the three-form multiplet and the standard 
scalar multipelt. 

The field strength \eqref{2.4} and 
the action \eqref{2.3} are invariant under gauge transformations
\bea
\d V = L~, \qquad \bar L =L~, \qquad \bar D^2 L=0~,
\label{2.6}
\eea
where $L$ is an arbitrary linear multiplet.  This gauge symmetry 
corresponds to 
a gauge theory with linearly dependent generators, 
following the terminology of the Batalin-Vilkovisky quantisation \cite{BV}, and
therefore the quantisation of \eqref{2.3} cannot be carried out using the 
Faddeev-Popov procedure.\footnote{Covariant 
quantisation of models for the three-form multiplet coupled to supergravity 
was carried out in \cite{BK88} (see \cite{BK} for a review).}

The supersymmetric theories \eqref{2.1} and \eqref{2.3}
describe the dynamics of two different multiplets,
the vector and the scalar ones, respectively, 
if the dynamical superfield $V$ is unconstrained. 
However, in case $V$ is subject to the nilpotency constraints \eqref{1.1},
which are incompatible with the gauge symmetries \eqref{2.2} and \eqref{2.6},
any difference between the actions  \eqref{2.1} and \eqref{2.3} proves to disappear. 
Once the constraints \eqref{1.1} are taken into account,
the  action \eqref{2.1}
(which coincides with the right-hand side of \eqref{1.4} modulo a total derivative)
can be rewritten in the form \eqref{2.3}. 
As a consequence of \eqref{1.1}, the chiral scalar $\J $
proves to be nilpotent, 
\begin{subequations} \label{2.7}
\bea
\J^2 =0\quad \Longrightarrow \quad \J = -\frac{D^\a \J D_\a \J}{D^2 \J}~.
\label{2.7a}
\eea
It also obeys the nonlinear constraint
\bea
\J = 2 \bar D^2 \frac{\bar \J \J}{D^2  \J +\bar D^2 \bar \J} ~,
\label{2.7b}
\eea
\end{subequations}
which expresses the field strength of the gauge three-form, ${\rm Im} F$, in terms of the Goldstino 
and ${\rm Re} F$.
The latter constraint follows from the observation that 
\eqref{1.3} may be rewritten in several equivalent forms 
\bea
V = -4 \frac{\bar \J \J}{\bar D^2 \bar \J } = -4 \frac{\bar \J \J}{D^2  \J }
 = -8 \frac{\bar \J \J}{D^2 \J + \bar D^2 \bar \J } ~.
\eea

The above analysis demonstrates that the Goldstino multiplet $V$
under the nilpotency conditions \eqref{1.1} is equivalent to
the  chiral superfield $\J$  subject to the constraints \eqref{2.7}.
Therefore, the description in terms of $\Psi$ provides a different realisation of the same multiplet.

It is also worth pointing out that the chiral superfield $\f$ defined by \eqref{1.8}
is the three-form multiplet associated with the irreducible 
Goldstino superfield $\cV$ defined by the constraints \eqref{1.1} and \eqref{1.6}, 
\bea
\f = -\frac{1}{4} \bar D^2 \cV ~.
\eea


\section{Irreducible and reducible Goldstino superfields}
\label{section3}

In accordance with the discussion in \cite{BHKMS}, there are two general types of  
$\cN=1$ Goldstino superfields, irreducible and reducible ones. 
Every irreducible Goldstino superfield contains only one independent
component field -- the Goldstino itself, while the other component fields are composites constructed from the Goldstino.\footnote{All known irreducible scalar Goldstino superfields are nilpotent, and the important examples are provided 
by eqs. \eqref{1.8a} and \eqref{1.10b}.} 
Given an irreducible Goldstino superfield, the  corresponding component
action can be related to the Volkov-Akulov action \cite{VA,AV} using the formalism developed in \cite{KT0}.
Every reducible Goldstino superfield 
contains at least two independent fields, one of which is the Goldstino and the other
fields are auxiliary (the latter become descendants of the Goldstino on the mass shell).

Examples of irreducible Goldstino superfields are the chiral scalar \eqref{1.8} and 
the complex linear scalar \eqref{cls}. The well-known example of a reducible Goldstino superfield was introduced in \cite{Casalbuoni,KS}.
It is a chiral scalar  $\cX$, $\bar D_\ad \cX =0$, 
which is subject to the nilpotency constraint
\bea
\qquad \cX^2 =0~,
\label{3X.1}
\eea
in conjunction with the requirement that  the descendant 
$D^2 \cX $ is nowhere vanishing.
The dynamics of this supermultiplet is described by the action 
\bea
S = \int \rd^4 x \rd^2 \q  \rd^2 \bar{\q} \, \bar \cX \cX 
-  \Big\{f  \int \rd^4 x \rd^2 \q \, \cX + {\rm c.c.} \Big\}~.
\label{3X.2}
\eea
As was pointed out in \cite{BHKMS}, 
the nilpotency condition \eqref{3X.1} is preserved if $\cX$ is locally rescaled, 
\bea
\cX ~\to ~\re^{\t} \cX~, \qquad \bar D_\ad \t=0~.
\label{3X.3}
\eea
Requiring the action \eqref{3X.2} to be stationary under such re-scalings  of $\cX$ 
(see \cite{KMcAT-M})
gives 
\bea
\cX = \f ~, 
\eea
where $\f$ is Ro\v{c}ek's chiral scalar defined by \eqref{1.8}. 

As was demonstrated in \cite{KTyler} (see also \cite{BHKMS,KM}), every irreducible Goldstino superfield 
can be expressed in terms of the complex linear Goldstino superfield $\S$, 
eq.  \eqref{cls}, 
and its conjugate $\bar \S$. 
On the other hand, $\S$ can be represented as a descendant of the reducible Goldstino superfield $V$ subject to the constraints \eqref{1.1}. 
The corresponding  realisation is 
\bea
\S = -4 f \frac{D^2 V}{\bar D^2 D^2 V} = - 4 f \frac{\bar \J}{\bar D^2 \bar \J}~.
\label{3.1}
\eea
A remarkable feature of this representation is that $\S$ proves to be invariant under 
 local re-scalings  \eqref{1.5} of $V$, 
\bea
\d_\t V = \t V \quad \Longrightarrow \quad \d_\t \S =0~.
\label{3.2}
\eea
Here $\t$ is an arbitrary real scalar superfield. Since every irreducible Goldstino superfield is a descendant of $\S$ and $\bar \S$,  we conclude that 
all irreducible Goldstino superfields are invariant under local re-scalings  \eqref{1.5} of $V$. This includes the real Goldstino superfield $\cV$ given by \eqref{1.9}, with 
$\S$ realised as in \eqref{3.1}.  

Now let us note that a transformation~\eqref{1.5}, where $V$ is subject to~\eqref{1.1}  rescales the auxiliary field.
Hence, the invariance of $\Sigma$ under this transformation means that $\Sigma$ does not contain the auxiliary field and describes
an irreducible Goldstino model. This observation also explains why all irreducible Goldstino models (which contain only the Goldstino 
and no auxiliary fields) are invariant under re-scaling of $V$. 

As was noticed in \cite{BHKMS}, every reducible Goldstino superfield can always be represented as an irreducible one plus a ``matter'' superfield, which contains
all the auxiliary component fields. In our case, the reducible Goldstino superfield $V$
can be realised as 
\bea
V = \cV + U~,
\qquad \cV= \frac{1}{f} \bar \S \S ~,
\eea
where $\S$ is given by \eqref{3.1}. The  ``matter'' superfield $U$ obeys the generalised nilpotency condition
\bea
U^2 + 2 \cV U =0
\eea
and transforms under \eqref{3.2} as 
\bea
\d_\t U = \t (\cV + U )~.
\eea
The superfield $U$ contains only one independent component field, which is the auxiliary field of $V$.

It should be remarked that $\S$ can also be expressed in terms of $\bar \cX$ as
\bea
\S =  - 4 f \frac{\bar \cX}{\bar D^2 \bar \cX}~,
\eea
compare with \eqref{3.1}. One may check that $\S$ is invariant under 
arbitrary re-scalings \eqref{3X.3}. This implies that all irreducible Goldstino superfields, 
realised as descendants of $\cX$ and $\bar \cX$,  are invariant under local re-scalings  \eqref{3X.3} of $\cX$. 


\section{Nilpotent three-form Goldstino multiplet}
\label{section4}


In this section, we introduce a new reducible Goldstino superfield. 
It is a three-form multiplet, 
\begin{subequations}\label{zh}
\bea
\cY = -\frac{1}{4} \bar D^2 \cU ~,\qquad \bar \cU =\cU~,
\label{zh1}
\eea
which is constrained to be nilpotent,
\bea
\cY^2=0~,
\label{zh2}
\eea 
\end{subequations}
in conjunction with the requirement that  the descendant 
$D^2 \cY$ is nowhere vanishing. The relations \eqref{zh} are invariant under 
gauge transformations of the type \eqref{2.6}.
The dynamics of $\cY$ is described by the action 
\bea
S= \int \rd^4 x \rd^2 \q  \rd^2 \bar{\q} \, \bar \cY \cY
-  \Big\{h  \int \rd^4 x \rd^2 \q \, \cY + {\rm c.c.} \Big\}~,
\label{zh3}
\eea
with $h=\bar h$ a positive parameter.  This nilpotent three-form multiplet 
is a variant formulation of the nilpotent chiral multiplet $\cX$ discussed in the previous section. 

Interesting enough, the above theory can be obtained from a nonlinear $\s$-model action 
\bea
S= \int \rd^4 x \rd^2 \q  \rd^2 \bar{\q} \, K( \bar \cY ,  \cY)  
+  \Big\{  \int \rd^4 x \rd^2 \q \, W(\cY) + {\rm c.c.} \Big\}
\label{zh4}
\eea
in the infrared limit, as an extension of the approach advocated in~\cite{Casalbuoni}. 
Let us define the components of $\cY$ as 
\be
\cY| =\varphi\,, \qquad D_{\a} \cY|= \sqrt{2} \psi_{\a}\,, \qquad -\frac{1}{4} D^2 \cY| = F = H +
\ri G~. 
\label{zh5}
\ee
As was already explained in eq.~\eqref{2.5} it follows from~\eqref{zh1} that $G$ is a divergence of a vector which we denote by $C^a$, that is 
$G = \partial_a C^a$. Equivalently,  $*G =d C$ for some three-form potential $C$. The action~\eqref{zh4} in components 
is just the standard action for a supersymmetric nonlinear $\s$-model 
\bea
S & = & \int \rd^4 x\, \Big[K_{\varphi \bar \varphi} (-\partial_a \varphi \partial^a \bar \varphi +\frac{\ri}{2} \partial_{\a \ad} \psi^{\a} \bar \psi^{\ad}
- \frac{\ri}{2}  \psi^{\a} \partial_{\a \ad}  \bar \psi^{\ad}  + F \bar F)
\nonumber \\
& + & \frac{1}{2} K_{\varphi \varphi \bar \varphi} ( \ri \psi^{\a} \bar \psi^{\ad} \partial_{\a \ad}\varphi  - \psi^2 \bar F) 
- \frac{1}{2} K_{\varphi \bar \varphi \bar \varphi} ( \ri \psi^{\a} \bar \psi^{\ad} \partial_{\a \ad} \bar \varphi + {\bar \psi}^2  F) 
+ \frac{1}{4} K_{\varphi \varphi \bar \varphi \bar \varphi} \psi^2 {\bar \psi}^2
\nonumber \\
&-& \frac{1}{2} W_{\varphi \varphi } \psi^2 -   \frac{1}{2} {\bar W}_{\bar \varphi \bar \varphi } {\bar \psi}^2
+W_{\varphi} F + {\bar W}_{\bar \varphi} \bar F \Big]~.
\label{zh6}
\eea
To achieve gauge invariant boundary conditions
\be 
\delta G|_{{\rm boundary}}=0
\label{zh6.5}
\ee
this action has to be supplemented with the boundary term\footnote{We will not discuss the supersymmetric completion of~\eqref{zh6.6}
in this note.} 
\cite{Duncan:1989ug, Groh:2012tf}
\be
S_{{\rm boundary}}= -\int \rd^4 x \,\partial_a [ 2 K_{\varphi \bar \varphi} G C^a + \ri (W_{\varphi} - {\bar W}_{\bar \varphi})C^a]~. 
\label{zh6.6}
\ee
The equations of motion for the auxiliary fields $H$ and $C^a$ are given by
\be
2 K_{\varphi \bar \varphi} H + W_{\varphi} + {\bar W}_{\bar \varphi}=0\,, \qquad 
\partial_a ( K_{\varphi \bar \varphi} G + \frac{\ri}{2} (W_{\varphi} - {\bar W}_{\bar \varphi}))=0 ~,
\label{zh8}
\ee
which lead to 
\be 
H= -\frac{1}{2} K_{\varphi \bar \varphi}^{-1} (W_{\varphi} + {\bar W}_{\bar \varphi})\,, \quad 
G= -\frac{\ri}{2} K_{\varphi \bar \varphi}^{-1} (W_{\varphi} -{\bar W}_{\bar \varphi})
+K_{\varphi \bar \varphi}^{-1} g~, 
\label{zh9}
\ee
where $g$ is an arbitrary constant. Substituting~\eqref{zh9} into the action~\eqref{zh6}, \eqref{zh6.6} we obtain the scalar potential
\be 
V( \varphi, \bar \varphi)= K_{\varphi \bar \varphi}^{-1} (W_{\varphi} + {\rm i} g) ( \bar W_{\bar \varphi} -{\rm i} g)~.
\label{zh10}
\ee
Note that it receives contributions from both the bulk and boundary actions. 
Let us choose K\"ahler normal coordinates in the $\s$-model target space
near the vacuum so that {\it in the vacuum}
\be 
K_{\varphi \bar \varphi} = 1\,, \quad K_{\varphi \varphi \bar \varphi} 
=K_{\varphi \bar \varphi \bar \varphi}=0\,, 
\quad K_{\varphi \varphi \bar \varphi \bar \varphi} =R~,
\label{zh7}
\ee
where $R$ is the curvature of the target space. Then the vacuum is determined by the equation
\be 
(\bar W_{\bar \varphi} - {\rm i} g)  W_{\varphi  \varphi}=0~. 
\label{z11}
\ee
We are interested in a supersymmetry breaking solution for which $\langle F \rangle \sim \langle \bar W_{\bar \varphi} - {\rm i} g \rangle \neq 0$. 
This means that $\langle W_{\varphi  \varphi} \rangle =0$. The vacuum energy is given by 
\be 
\Lambda  = \langle V( \varphi, \bar \varphi) \rangle = |W_{\varphi} +{\rm  i} g |^2 >0~.
\label{zh12}
\ee
Since $\langle W_{\varphi  \varphi} \rangle =0$, the fermion $\psi$ is massless and becomes a Goldstino. In addition, we have two massive scalar fields. 
For simplicity let us assume that the supersymmetry breaking vacuum is located at $\varphi =\bar \varphi=0$. Denoting 
\be 
W_{\varphi} (0)=- h =- \bar h\,, \quad \frac{1}{2} W_{\varphi \varphi \varphi} (0)= w= \bar w~,
\label{zh13}
\ee
where for simplicity we have assumed that $h$ and $w$ are real,
and expanding the action to quadratic order in fluctuations we find that the masses of the scalar fields are 
\be 
m_{\pm}^2= - |{\bm f}| (|{\bm f}| R \pm w)~,
\label{zh14}
\ee
where ${\bm f} = h + {\rm i} g$.
To make sure that they both are positive we have to require that $R <0$.  

To find the theory of just the Goldstino $\psi$ we decouple the scalar fields by taking their masses to infinity. For this we take the 
limit of  infinite curvature $|R| \to \infty$. The leading contribution to the action~\eqref{zh6} in this limit is 
\bea 
S_{{\rm div}} &= & R \int \rd^4 x \,\Big[ \varphi \bar \varphi 
(-\partial_a \varphi \partial^a \bar \varphi  + \frac{\ri}{2} \partial_{\a \ad} \psi^{\a} \bar \psi^{\ad}
- \frac{\ri}{2}  \psi^{\a} \partial_{\a \ad}  \bar \psi^{\ad}  + F \bar F ) 
\nonumber \\
& + & \frac{1}{2} \bar \varphi ( \ri \psi^{\a} \bar \psi^{\ad} \partial_{\a \ad}\varphi  - \psi^2 \bar F) 
- \frac{1}{2} \varphi ( \ri \psi^{\a} \bar \psi^{\ad} \partial_{\a \ad} \bar \varphi + {\bar \psi}^2 \bar F) 
+ \frac{1}{4} \psi^2 {\bar \psi}^2\Big]~. 
\label{zh15}
\eea
This gives the following equation of motion for $\phi$
\be 
\frac{1}{2} \bar \varphi \partial_a \partial^a \varphi^2
+ \bar F (\varphi F- \frac{1}{2} \psi^2) 
+ \ri \partial_{\a \ad} (\varphi \psi^{\a} \bar \psi^{\ad} ) - \ri \varphi \psi^{\a} \partial_{\a \ad} ( \bar \psi^{\ad} )=0~.
\label{zh16}
\ee
It is clear that this equation is solved by 
\be 
\varphi =\frac{1}{2 F} \psi^2~. 
\label{zh17}
\ee
Note that it implies that $\varphi^2 =\varphi \psi=0$. Eq.~\eqref{zh17} is precisely what follows from the nilpotency condition~\eqref{zh2}. 
Hence, we obtain that the theory of a nilpotent three-form multiplet specified by eqs.~\eqref{zh1}, \eqref{zh2} 
arises as the infrared limit of the $\sigma$-model~\eqref{zh4}. 
Since $S_{{\rm div}}=0$ on the solution~\eqref{zh17} the action for the remaining fields $\psi, H, G$ comes from the subleading terms in the action~\eqref{zh6}
\bea 
S[\psi, H, G] &= & \int \rd^4 x \,\Big[ -\partial_a \Big( \frac{\psi^2}{2F} \Big)\partial^a \Big( \frac{\bar \psi^2}{2\bar F} \Big)+\frac{\ri}{2} \partial_{\a \ad} \psi^{\a} \bar \psi^{\ad}
- \frac{\ri}{2}  \psi^{\a} \partial_{\a \ad}  \bar \psi^{\ad}  
\nonumber \\
&+& F \bar F - h (F +\bar F)  \Big]\,, \qquad F= H +\ri G~.
\label{zh18}
\eea
Note that this is just the component form of the superfield action~\eqref{zh3}.
This action should be supplemented by the boundary term
\be
S_{{\rm boundary}}[\psi, H, G] = - 2\int \rd^4 x \,\partial_a ( G C^a )~.
\label{zh19}
\ee
The auxiliary fields $H$ and $G$ can be eliminated using their equations of motion which gives
\begin{subequations} 
\bea
H - h - \frac{\bar \psi^2}{4{\bar F}^2} \Box \frac{\psi^2}{2F} - \frac{\psi^2}{4 F^2} \Box \frac{{\bar \psi}^2}{2\bar F} &=&0~, 
\label{zh20.1}
\\
\partial_a \Big[G  +  \ri \frac{\bar \psi^2}{4{\bar F}^2} \Box \frac{\psi^2}{2F} - \ri \frac{\psi^2}{4 F^2} \Box \frac{{\bar \psi}^2}{2\bar F} \Big]&=&0~.
 \label{zh20.2}
\eea
\end{subequations} 
Eq.~\eqref{zh20.2} is solved by 
\be 
G  + \frac{\ri}{2}  \Big( \frac{\bar \psi^2}{2{\bar F}^2} \Box \frac{\psi^2}{2F} - \frac{\psi^2}{2 F^2} \Box \frac{{\bar \psi}^2}{2\bar F} \Big) =g
\label{zh21}
\ee
for some constant $g$. 
Eqs.~\eqref{zh20.1} and~\eqref{zh21} can also be written as
\be
F - {\bm f}  - \frac{\bar \psi^2}{2{\bar F}^2} \Box \frac{\psi^2}{2F} =0~,
\label{zh22}
\ee
where, as before, ${\bm f} = h+ \ri g$.


\section{Models for spontaneously broken supergravity}

In describing supergravity-matter systems in superspace, 
it is useful, following the ideas  pioneered in  \cite{Siegel77_2,KakuT},
to formulate every pure supergravity theory
 as conformal supergravity coupled to a compensating supermultiplet. 
 Different off-shell formulations for  supergravity correspond to different compensators
 \cite{GGRS,Ferrara:1983dh}.
Conformal supergravity can be realised 
 using the supergravity multiplet described in terms of the 
 superspace geometry of \cite{GWZ} 
(which underlies the Wess-Zumino approach \cite{WZ} to the old minimal formulation for $\cN=1$ supergravity developed independently in \cite{old1,old2}) 
augmented with the super-Weyl invariance  of \cite{HT}.
An infinitesimal super-Weyl transformation has  the form
\begin{subequations} 
\label{superweyl}
\bea
\d_\s \cD_\a &=& ( {\bar \s} - \hf \s)  \cD_\a + (\cD^\b \s) \, M_{\a \b}  ~, \\
\d_\s \bar \cD_\ad & = & (  \s -  \hf {\bar \s})
\bar \cD_\ad +  ( \bar \cD^\bd  {\bar \s} )  {\bar M}_{\ad \bd} ~,\\
\d_\s \cD_{\a\ad} &=& \hf( \s +\bar \s) \cD_{\a\ad} 
+\frac{\ri}{2} (\bar \cD_\ad \bar \s) \cD_\a + \frac{\ri}{2} ( \cD_\a  \s) \bar \cD_\ad \non \\
&& + (\cD^\b{}_\ad \s) M_{\a\b} + (\cD_\a{}^\bd \bar \s) \bar M_{\ad \bd}~,
\eea
\end{subequations}
where $\s$ is an arbitrary covariantly chiral scalar superfield,  $\bar \cD_\ad \s =0$. 

There are several natural ways to lift the Goldstino superfield $V$, 
which was described in section 
\ref{section1}, to supergravity. Of course, the constraints \eqref{1.1} are generalised to 
curved superspace in the unique way given in  \cite{KMcAT-M,BHKMS}: 
\begin{subequations} \label{5.2}
\bea
V^2&=&0~, \\
V \cD_A \cD_B V &=&0~,\\
V \cD_A \cD_B \cD_C V &=&0~. 
\eea
\end{subequations}
Non-uniqueness occurs when choosing a super-Weyl transformation for $V$.
In this paper we consider the following transformation laws:
\bea
\d_\s V^{(\rm I)} &=& 0~, 
\label{5.3}
\eea
and 
\bea
\d_\s V^{(\rm II)}  &=& (\s+ \bar \s) V^{(\rm II)} ~.
\label{5.4}
\eea
In both cases, the constraints \eqref{5.2} are super-Weyl invariant. 
One can choose a more general super-Weyl transformation law for $V$ of the form 
$\d_\s V = a (\s +\bar \s) V$, for some parameter $a$, by multiplying 
$ V^{(\rm I)} $ by some power of the  compensator used. In particular, 
the choice $a=-1$ was made in \cite{BHKMS} to describe the Goldstino brane.


\subsection{Super-Weyl inert $V$}

Let us assign the super-Weyl transformation law \eqref{5.3} to the Goldstino superfield 
$V$. Then the coupling of the old minimal supergravity \cite{WZ,old1,old2}
to $V$ is described by 
the super-Weyl invariant action \cite{KMcAT-M}
\bea
S &=& - \frac{3}{ \k^2}
\int {\rm d}^{4} x \rd^2\q\rd^2\bar\q\,
E\,  \bar S_0 S_0  
+ \Big\{ \frac{ \m}{ \k^2}   \int {\rm d}^{4} x \rd^2 \q\,
{\cal E} \,S_0^3 
+ {\rm c.c.} \Big\} \non \\
&&  +\int \rd^4 x \rd^2 \q  \rd^2 \bar{\q} \, E\,\Big\{
\frac{1}{16} V \cD^\a (\bar \cD^2 -4R ) \cD_\a V- 2f \bar S_0 S_0 V\Big\}~.
\label{5.5}
\eea
Here $S_0$ the chiral compensator, $\bar \cD_\ad S_0 =0$, 
with the super-Weyl transformation $\d_\s S_0 = \s S_0$.\footnote{Here 
$E$ is the full superspace measure,  
and  $\cE$ denotes the chiral density. 
We use the notation $S_0$ for the chiral compensator
following \cite{KU,Ferrara:1983dh}. In the superspace literature,  
the chiral compensator is usually denoted $\F$, see e.g. section 6.6.1 in \cite{BK}. } 
The functional in the first line of \eqref{5.5} is the  old minimal supergravity action,
in which the expression in the figure brackets 
is the supersymmetric cosmological term \cite{FvN2,Siegel78}. 

Within the new minimal formulation for $\cN=1$ supergravity 
\cite{new} (see also \cite{SohniusW2,SohniusW3}), 
the compensator is a  real covariantly linear scalar superfield,
\bea  
(\bar \cD^2 -4R) {\mathbb L}  =0~, \qquad \bar {\mathbb L}= {\mathbb L}~,
\eea
with the super-Weyl transformation $\d_\s {\mathbb L} = (\s+\bar \s) {\mathbb L}$.
The action for supergravity coupled to the Goldstino superfield $V$ is \cite{KMcAT-M}
\bea
S=   \int \rd^4 x \rd^2 \q  \rd^2 \bar{\q} \, E\, \left\{\frac{3}{\k^2}
{\Bbb L}\, {\rm ln} \frac{\Bbb L}{|S_0|^2}
+\frac{1}{16} V \cD^\a (\bar \cD^2 -4R ) \cD_\a V- 2f {\mathbb L} V
 \right\}~,
 \label{5.7}
\eea
where  $S_0$ is a covariantly chiral scalar superfield, $\bar \cD_\ad S_0=0$ 
(it is a purely gauge degree of freedom). The first term in \eqref{5.7} corresponds
to the new minimal  supergravity action. An important feature of unbroken new minimal supergravity is that it does not allow 
any supersymmetric cosmological term \cite{SohniusW2,SohniusW3}.\footnote{Among the known off-shell 
formulations for $\cN=1$ supergravity (see \cite{GGRS,BK} for reviews), supersymmetric cosmological terms 
exist only for the old minimal supergravity \cite{FvN2,Siegel78} 
(and its variant version reviewed in section \ref{subsection5.3} below)
and the $n=-1$ non-minimal supergravity of \cite{ButterK12}.}


\subsection{Super-Weyl non-invariant $V$}

If the Goldstino superfield $V$ is chosen to possess the super-Weyl transformation law \eqref{5.4}, then 
we can associate with it the following three-form superfield
\bea
\J = -\frac{1 }{ 4} ({\bar \cD}^2 -4R) V~.
\eea
It is nilpotent, $ \J^2 =0$, as a consequence of \eqref{5.2}.
It transforms as a primary superfield under the super-Weyl group, 
\bea
\d_\s \J = 3\s \J~.
\eea
The action for old minimal supergravity coupled to the Goldstino superfield $\J$  is 
\bea
S &=& 
\int {\rm d}^{4} x \rd^2\q\rd^2\bar\q\,
E\, \Big(- \frac{3}{ \k^2} \bar S_0 S_0 + \frac{\bar \J  \J}{(\bar S_0 S_0)^2}
 \Big) \non \\
&& \qquad \qquad \qquad 
+ \left\{  \int {\rm d}^{4} x \rd^2 \q\,
{\cal E} \, \Big(  \frac{ \m}{ \k^2}S_0^3 -f \J  \Big)
+ {\rm c.c.} \right\}
~.
\eea
This model is equivalent to \eqref{5.5}.


\subsection{Three-form supergravity}
\label{subsection5.3}

In 1981, Gates and Siegel \cite{GS} proposed a variant formulation of the old minimal supergravity, which is obtained from the supergravity action given in \cite{SG79} by replacing the compensating chiral scalar  superfield  with a three-form 
superfield.\footnote{Probably the first serious study of matter three-form multiplets coupled
to supergravity was given in \cite{BK88}. A superform formulation for the three-form 
multiplet in a supergravity background was given in \cite{BPGG}.}
At the component level, this supergravity formulation was elaborated in \cite{OW}
where is was called ``three-form supergravity.''
The modern prepotential formulation for three-form supergravity was given 
in Appendix B of \cite{KMcC}. 
Further aspects of three-form supergravity were studied in \cite{BM}.

Here we describe three-form supergravity following \cite{KMcC}.
This supergravity formulation is obtained from the old minimal theory by replacing 
 $S_0^3$ with
\bea
\X^3 = -\frac{1 }{ 4} ({\bar \cD}^2 -4R) P~,
\qquad {\bar P} =P~, 
\label{5.11}
\eea
where the prepotential $P$ is a real scalar superfield such that $\X$ is nowhere
vanishing.
The super-Weyl  transformation of $P$ is defined to be
\bea
\d_\s P =( \s +{\bar \s} )P~.
\label{s-weyl-P}
\eea
Then the  chiral superfield $\X$ transforms as 
\bea
\d_\s \X = \s \X~.
\eea
The pure supergravity action is 
\bea
S_{\rm SG} &=& - \frac{3}{\k^2}
 \int \rd^4 x \rd^2 \q  \rd^2 \bar{\q} \,E\, {\bar \X}  \X
+\frac{2m}{\k^2}  \int \rd^4 x \rd^2 \q  \rd^2 \bar{\q} \,E\, P 
\non \\
&=& -\frac{3}{\k^2}
 \int \rd^4 x \rd^2 \q  \rd^2 \bar{\q} \,E\,  {\bar \X}  \X  
+\frac{m}{\k^2}  \Big\{   \int \rd^4 x \rd^2 \q \,  \cE  \,
\X^3 +{\rm c.c.} \Big\}~. 
\label{omsg3}
\eea
Here the second term on the right is a
supersymmetric cosmological term. 

The supergravity action (\ref{omsg3})
is invariant  under gauge transformations 
\be 
\d P = L~, \qquad 
({\bar \cD}^2 - 4R ) \, L = 0 ~,
\label{5.15}
\ee 
which  leave invariant the chiral superfield $\X^3$ defined by \eqref{5.11}.
The equation of motion for $P$ is 
\bea
{\mathbb R} + \bar {\mathbb R} =2m~, 
\qquad {\mathbb R} := -\frac{1}{4} \X^{-2} (\bar \cD^2 - 4R) \bar \X~.
\eea
Here the chiral scalar $\mathbb R$ is invariant under the super-Weyl transformations.
The general solution to the equation of motion is ${\mathbb R}  = m +\ri e$, for some real 
parameter $e$.


\subsection{Three-form Goldstino superfield in supergravity}

The nilpotent three-form multiplet \eqref{zh}
is naturally lifted to supergravity. In curved superspace, 
its gauge-invariant field strength is defined by
\begin{subequations}
\label{5.17}
\bea
\cY = -\frac{1}{4} (\bar \cD^2 -4R) \cU ~,\qquad \bar \cU =\cU~.
\label{5.17a}
\eea
This chiral scalar is constrained to be nilpotent,
\bea
\cY^2=0~,
\label{5.17b}
\eea 
\end{subequations}
in conjunction with the requirement that  the descendant 
$\cD^2 \cY$ is nowhere vanishing.
The prepotential $\cU$ in \eqref{5.17a} is postulated to possess the super-Weyl 
transformation 
\bea
\d_\s \cU =( \s +{\bar \s} )\cU~,
\label{5.18}
\eea
which implies that $\cY$ is transforms as a super-Weyl primary superfield, 
\bea
\d_\s \cY = 3\s \cY~.
\eea
The chiral scalar $\cY$ defined by \eqref{5.17a} is invariant under gauge transformations
of $U$ of the type \eqref{5.15}.

The action for old minimal supergravity coupled to the Goldstino superfield $\cY$  is 
\bea
S &=& 
\int {\rm d}^{4} x \rd^2\q\rd^2\bar\q\,
E\, \Big(- \frac{3}{ \k^2} \bar S_0 S_0 + \frac{\bar \cY  \cY}{(\bar S_0 S_0)^2}
 \Big) \non \\
&& \qquad \qquad \qquad 
+ \left\{  \int {\rm d}^{4} x \rd^2 \q\,
{\cal E} \, \Big(  \frac{ \m}{ \k^2}S_0^3 -h \cY  \Big)
+ {\rm c.c.} \right\}
~.
\label{5.20}
\eea
In the flat superspace limit it reduces to \eqref{zh3}.
This action should be supplemented by a boundary term
generalising \eqref{zh19}.

As discussed in section \ref{section4}, the Goldstino superfield $\cY$ contains two independent
auxiliary fields, $F= H+\ri G$, 
of which $H$  is a scalar and $G$ is the field strength of a gauge three-form. 
In supergravity, both $H$ and $G$ produce positive contributions to the cosmological constant.  While the contribution from $H$ is uniquely determined by the parameter of the supersymmetry breaking $h$ in \eqref{5.20}, 
the contribution from $G$ is dynamical. 
We believe that the latter may be used to neutralise the negative contribution from the supersymmetric cosmological term.

The idea that  massless gauge three-forms make it possible to generate a cosmological constant dynamically, has attracted much interest since the early  1980s 
\cite{OS,DvN,ANT,Hawking,Duff,Duncan:1989ug,BP}. We believe it will be useful in the framework of spontaneously broken local supersymmetry. 

After this work had been submitted to the hep-th archive, we became aware 
of the recent paper \cite{Farakos:2016hly} in which the nilpotent three-form multiplet
was also studied and used for different purposes.  
\\

\noindent
{\bf Acknowledgements:}\\
E.I.B. would like to thank the Physics Department at the University of Pennsylvania
where some of this work was done for warm hospitality.
The work of E.I.B. was  supported in part by the ARC Future Fellowship FT120100466.
Both authors were  supported in part by the Australian 
Research Council, project No. DP140103925.
The work of SMK was also supported in part by the Australian 
Research Council, project No. DP160103633.

\begin{footnotesize}

\end{footnotesize}

\end{document}